\documentclass[a4paper]{article} 

\def\comment#1{} 
\def\journalfont{\rm}         
\def\jou#1{{\journalfont #1\ }}
\def\joudef#1#2{\def #1{\jou{\ignorespaces #2}}}

\joudef{\aaa}    { Astron.\ Astrophys.}
\joudef{\aip}    { Adv.\ Phys.}
\joudef{\adm}    { adv.\ math.}
\joudef{\am}     { Ann.\ Math.}
\joudef{\apl}    { Ann.\ Phys.\ (Leipzig)}
\joudef{\apny}   { Ann.\ Phys.\ (N.Y.)}
\joudef{\arnps}  { Annu.\ Rev.\ Nucl.\ Part.\ Sci.}
\joudef{\apj}    { Astrophys.\ J.}
\joudef{\apjl}    { Astrophys.\ J.\ Lett.}
\joudef{\cjp}    { Can.\ J.\ Phys.}
\joudef{\cmp}    { Commun.\ Math.\ Phys.}
\joudef{\cqg}    { Class.\ Quantum Grav.}
\joudef{\grg}    { Gen.\ Rel.\ Grav.}
\joudef{\ijmpd}  { Int.\ J.\ Mod.\ Phys.\ D}
\joudef{\ijtp}   { Int.\ J.\ Theor.\ Phys.}
\joudef{\invm}   { Invent.\ Math.}
\joudef{\jm}     { J.\ Math.}
\joudef{\jmaa}   { J.\ Math.\ Anal.\ Appl.}
\joudef{\jmp}    { J.\ Math.\ Phys.}
\joudef{\jpa}    { J.\ Phys.\ A}
\joudef{\lr}    { Liv.\ Rev.\ Rel.}
\joudef{\mnras}  { Mon.\ Not.\ R.\ Ast.\ Soc.}
\joudef{\mpl}   { Mod.\ Phys.\ Lett.} 
\joudef{\mpla}   { Mod.\ Phys.\ Lett.\ A} 
\joudef{\nature} { Nature}
\joudef{\nc}     { Nuovo Cim.}
\joudef{\npb}    { Nuc.\ Phys.\ B}
\joudef{\ph}     { Physica}
\joudef{\pla}    { Phys.\ Lett. A}
\joudef{\plb}    { Phys.\ Lett. B}
\joudef{\pr}     { Phys.\ Rev.}
\joudef{\pra}    { Phys.\ Rev.\ A}
\joudef{\prb}    { Phys.\ Rev.\ B}
\joudef{\prc}    { Phys.\ Rev.\ C}
\joudef{\prd}    { Phys.\ Rev.\ D}
\joudef{\prep}   { Phys.\ Rep.}
\joudef{\prl}    { Phys.\ Rev.\ Lett.}
\joudef{\prsla}  { Proc.\ Roy.\ Soc.\ Lond.\ A}
\joudef{\ptp}    { Prog.\ Theor.\ Phys.}
\joudef{\ptps}   { Prog.\ Theor.\ Phys.\ Suppl.}
\joudef\rmp      { Rev.\ Mod.\ Phys.}
\joudef\spj      { Sov.\ Phys.\ JETP}

\catcode`@=11

\def\eqalign#1{\null\,\vcenter{\openup\jot\m@th
  \ialign{\strut\hfil$\displaystyle{##}$&$\displaystyle{{}##}$\hfil
      \crcr#1\crcr}}\,}
\def\meqalign#1{\null\,\vcenter{\openup\jot\m@th
  \ialign{\strut\hfil$\displaystyle{##}$&&$\displaystyle{{}##}$\hfil
      \crcr#1\crcr}}\,}
\catcode`@=12   

\newdimen\arrayruleHwidth
\makeatletter
\def\Hline{\noalign{\ifnum0=`}\fi\hrule \@height \arrayruleHwidth
  \futurelet \@tempa\@xhline}
\makeatother

\newcommand\thickbaselines{\baselineskip=20pt\lineskip=3pt\lineskiplimit=3pt}
\catcode`@=11
\def\cases#1{\left\{\,\vcenter{\thicknormalbaselines\m@th
             \ialign{$##\hfil$&\quad##\hfil\crcr#1\crcr}}\right.}
\def\matrix#1{\null\,\vcenter{\thickbaselines\m@th
    \ialign{\hfil$##$\hfil&&\quad\hfil$##$\hfil\crcr
      \mathstrut\crcr\noalign{\kern-\baselineskip}
      #1\crcr\mathstrut\crcr\noalign{\kern-\baselineskip}}}\,} 
\catcode`@=12   

\newcommand{\eprint}{\textsf} 

\newcommand\be{\begin{equation}} \newcommand\ee{\end{equation}} 
\newcommand\bd{\begin{displaymath}}\newcommand\ed{\end{displaymath}}

\renewcommand{\d}{{\rm d}} 
\newcommand{\e}{{\rm e}}

\newcommand\Tr{\mathop{\rm Tr}\nolimits}
\newcommand\ts\textstyle

\def\undersim#1{\mathop{\vtop{\ialign{##\crcr
     $\hfil\displaystyle{#1}\hfil$\crcr\noalign
     {\kern1pt\nointerlineskip}\hbox{$\hfil\sim\hfil$}\crcr
     \noalign{\kern1pt}}}}}

\newcommand{\acronym}[3]{\newcommand{#1}{#3 (#2)\relax\renewcommand{#1}{#2}}}

\def\eg{{\it e.g.}}  \def\ie{{\it i.e.}}

\newtoks\reportnoregister \newtoks\eprintnoregister
\newcommand{\reportnumber}[1]{\reportnoregister={#1}}
\newcommand{\eprintnumber}[1]{\eprintnoregister={#1}}

\reportnumber{\mbox{}} 
\eprintnumber{\mbox{}} 

\newcommand{\reportid}{
   \begin{minipage}{17cm}\vspace{-3.2cm}
     \begin{flushright}
      {\normalsize \the\reportnoregister \\[-.2cm]
            \eprint{\the\eprintnoregister}}\vspace{3.2cm}
     \end{flushright}
   \end{minipage}\hspace{-17cm} }

\catcode`@=11   
\def\title#1{\gdef\@title{\reportid#1}}
\catcode`@=12   

\acronym{\SSS}{SSS}{{\em static spherically symmetric}}
\acronym{\TOV}{TOV}{{\em Tolman-Oppenheimer-Volkoff}}

\newcommand{\ncd}{\newcommand}
\ncd{\nms}{\negmedspace}
\ncd{\nts}{\negthickspace}
\ncd{\mcl}[1]{\mathcal{#1}}
\ncd{\beq} {\begin{equation}}
\ncd{\eeq} {\end{equation}}
\ncd{\BE} {\begin{eqnarray}}
\ncd{\EE} {\end{eqnarray}}
\ncd{\rarr} {\rightarrow}
\ncd{\larr} {\leftarrow}
\ncd{\lrarr} {\leftrightarrow}
\ncd{\lbeq}[1]  {\label{eq: #1}}
\ncd{\refeq}[1] {(\ref{eq: #1})}
\ncd{\mrm}    {\mathrm}
\ncd{\nn}{\nonumber}
\ncd{\mbf}[1] {{\mathbf #1}}
\ncd\T{\frac{1}{2}h^{\mu\nu}p_\mu p_\nu}
\ncd{\ms}{\mathstyle}
\ncd{\ds}{\displaystyle}
\ncd{\Yc}{Y_{\rm c}}
\ncd{\Yorb}{Y_{\rm orb}}

\ncd{\der}{{\mathrm{d}}}
\ncd{\rtil}{\tilde{r}}
\ncd{\Mr}{\frac{2M}{r}}
\ncd{\rhotil}{\tilde{\rho}}
\ncd{\rstar}{r_{*}}

\usepackage{amsmath}
\usepackage{graphicx}
\usepackage{amssymb}

\ncd{\dell}{\partial}
\ncd{\mnote}[1]{\marginpar{\small #1}}

\ncd{\ec}{\check\epsilon}
\ncd{\rhoc}{\check\rho}
\ncd{\muc}{\check\mu}
\ncd{\pc}{\check{p}}
\ncd{\Gc}{\check\Gamma}
\ncd{\Oc}{\check\Omega}
\ncd{\betac}{\check\beta}

\ncd{\bldeta}{\boldsymbol{\eta}}
\ncd{\bldone}{\mathbf{1}}
\ncd{\blds}{\mathbf{s}}
\ncd{\bldk}{\mathbf{k}}
\ncd{\blde}{\mathbf{e}}

\ncd{\abs}[1] {|#1|} 
\ncd{\ubold}{\mathbf u}
\ncd{\Abold}{\mathbf A}
\ncd{\Bbold}{\mathbf B}
\ncd{\Mbold}{\mathbf M}
\ncd{\tsfrac}[2]{{\ts\frac{#1}{#2}}}
\ncd{\lagom}{\hspace{.6pt}}
\ncd{\muk}{k}
\ncd{\lagomdot}{{\mbox{\large$\cdot$}}}

\ncd{\stil}{\tilde{s}}
\ncd{\ftil}{\tilde{f}}
\ncd{\Otil}{\tilde{\Omega}}

\ncd{\D}{\mathcal{D}}
\ncd{\Qcal}{\mathcal{Q}}
\ncd{\Jcal}{\mathcal{J}}
\ncd{\Ecal}{\mathcal{E}}
\ncd{\Wcal}{\mathcal{W}}
\ncd{\Xcal}{\mathcal{X}}
\ncd{\vt}{v_{t\perp t}^{\,2}}
\ncd{\vr}{v_{r\perp}^{\,2}}

\ncd{\psimap}{\Psi}
\ncd{\psivar}{\psi}

\ncd{\Ap}{(r\psivar)'}
\ncd{\Adot}{(r\psivar)^{\lagomdot}}
\ncd{\Bp}{(r^{-1}\varphi)'}
\ncd{\Bdot}{(r^{-1}\varphi)^{\lagomdot}}

\ncd{\ela}{\left(1-\frac{2m}{r}\right)}
\ncd{\nfrac}[2]{\left(\frac{n_{#1}}{n_{#2}}\right)^2}

\ncd{\shm}{S}
\ncd{\shmtwoD}{\mathcal{\shm}}
\ncd{\lie}{\mathcal{L}}
\ncd{\brk}{\mathrm{max}}
\ncd{\fgauge}{f_\mathrm{G}}

\begin{document}

\title{Elastic stars in general relativity: IV. Axial perturbations}
\author{Max Karlovini$^1$\footnote{E-mail: \eprint{max@physics.muni.cz}}~~and 
        Lars Samuelsson$^2$\footnote{E-mail: \eprint{lars@soton.ac.uk}} \\[-10pt]
{\small $^1$ Department of Theoretical Physics and Astrophysics, Masaryk University,}\\[-15pt]
{\small Kotlarska 2, 611 37 Brno, Czech Republic} \\[-10pt]
{\small $^2$ School of Mathematics, University of Southampton, Southampton SO17~1BJ, UK} \\
\begin{minipage}[t]{0.8\linewidth}\small{ This is the fourth paper in a series that attempt 
to put forward a consistent framework for modelling solid regions in
neutron stars. Here we turn our attention to axial perturbations of
spherically symmetric spacetimes using a gauge invariant approach due
to one of us. Using the formalism developed in the first paper in the
series it turns out that the matter perturbations are neatly
expressible in terms of a ``metric'' tensor field depending only on
the speeds of shear wave propagation along the principal directions in
the solid. The results are applicable to a wide class of elastic
materials and does not assume material isotropy nor quasi-Hookean
behaviour. The perturbation equations are then specialised to a static
background and are given by two coupled wave equations. Our formalism
is thus slightly simpler than the previously existing results of
Shumaker \& Thorne, where an additional initial value equation needs
to be solved. The simplification is mainly due to the gauge invariance
of our approach and shows up also in somewhat simpler boundary
conditions. We also give a first order formulation suitable for
numerical integration of the quasi-normal mode problem of a neutron
star. The relations between the gauge independent variables and the,
in general, gauge dependent perturbed metric and strain tensor are
explicitly given.  }\end{minipage}}

\date{}

\maketitle

\section{Introduction}

\subsection{Motivation}

In recent years it has become increasingly clear that, in order to
extract information from observations of neutron stars, we must first
understand the dynamical behaviour of matter beyond the perfect fluid
approximation. Much effort have been put in to understand \eg\ the
dynamics of superfluids including various viscous phenomena, see
\cite{ac:sfreview} for a recent review. This topic is essential in
order to realistically discuss \eg\ glitches and gravitational waves
arising from secular instabilities. These events may also require
understanding of magnetic interactions if the field is strong enough
to influence the dynamics in the region where the matter currents are
strong. This is certainly the case at least in the outer layers of
magnetars and the outermost crust in pulsars but may even be important
in the interior regions if the magnetic field is buried during the
formation of the star. Glitches will require a thorough understanding
of the behaviour of the rigid parts of the star and its interactions
with normal and super-fluids. Recently a formalism was presented
\cite{cs:superelastic} suited for handling elastic solids
permeated by a superfluids, including, in a MHD like manner, magnetic
fields.

The objective of this series of papers is to develop a coherent
framework for purely elastic components in compact objects. In this,
the fourth paper, we consider axial perturbations of spherical
stars. It is worth pointing out that axial oscillations in neutron
stars may already have been observed in the aftermath of giant flares
in soft gamma ray repeaters, where quasi periodic oscillations with
frequencies from 18 Hz reaching into the kHz range are observed
\cite{israel:qpo,sw:qpo,sw:flare2,barat:qpo1979,sw:flare3}. 
Although the enormous magnetic fields believed to be present in these
objects, which are well modelled within the so-called magnetar model
\cite{dt:magnetars}, will couple any torsional oscillations to the
core within about one oscillation period via Alfv\'en waves
\cite{gsa:mhd,levin:magnetars} a recent toy-model suggests
\cite{gsa:mhd} that the frequencies will nevertheless be close to the
purely elastic ones. Therefore, even in this type of systems, it is
important to understand the elastic oscillation spectrum. This
spectrum has recently been studied, within the relativistic Cowling
approximation, in
\cite{sa:axicowling}. The local magnetic effects (\ie\ neglecting the 
global nature of the modes) of the torsional mode spectrum has been
studied for a poloidal dipole field \cite{sks:torsional} (see also
\cite{mps:qpo}). These calculations were recently extended to include
purely axial global oscillations \cite{sksv:torsionalII}.

The Newtonian theory of axial perturbations of elastic bodies is well
developed, see \eg\ \cite{mvh:nonrad}. The general relativistic
problem was pioneered by Schumaker and Thorne \cite{st:torsional}
(hereafter ST) who developed a detailed theory including many relevant
limits. Our treatment is different from theirs in several important
ways. First, we use the gauge invariant perturbation formalism due to
one of us \cite{karlovini:axial}. This formalism assumes that the
perturbations are axisymmetric, but as discussed \eg\ in
\cite{chandra:mathbh}, this implies no restriction for spherically
symmetric backgrounds. At a practical level, the difference
between our and ST's approach shows up in the fact that our final
equations consist of just two coupled wave equations and thereby
dispenses of the additional initial value equation, being traceable to
the gauge choice, in ST. Moreover, the formalism applies to any
spherically symmetric (static or not) background. More importantly,
our treatment does not from the outset assume an isotropic background
which makes it applicable to the exotic (``nuclear pasta'') phases
believed to exist in the lower crust of the neutron star (\eg\
\cite{pr:crustrev}).  Another minor difference is that our formalism
in the main part of this work is valid for non-linear elastic
equations of state. That is, we do not assume a quasi-Hookean equation
of state from the outset. Practically this should not be very
important since the behaviour of perfect elasticity (which we do
assume) is likely to break down roughly when any nonlinear
corrections to quasi-Hookean behaviour become important. 

As mentioned above, our treatment will not be limited to isotropic
backgrounds. There are two fundamentally different cases in which
non-isotropy occur, either the material is intrinsically non-isotropic
(as in the pasta phases) or the star is already strained. One may
think that the latter is fairly unimportant since, at any given time,
the crust will be close to its unstrained state simply because it will
break otherwise. However, the same limitations that apply to the
background applies to the perturbations so that the background strain
may well be of the same order as (or larger than!) the
oscillations. An example where this may be important is provided by
the flare model of Duncan \cite{duncan:1998A} where the initial crust quake sends
seismic waves in the (already strained) crust thereby causing it to
fracture at other locations too.

The treatment relies heavily on Paper I in the
series\cite{ks:relasticityI} (hereafter Paper I), where the
theoretical foundations of relativistic elasticity were outlined
following the framework of Carter and Quintana\cite{cq:elastica}. For
convenience and in order to introduce notation a brief review of the
formalism of Paper I is given below.

\subsection{Relasticity}

Any description of continuous media is, at least implicitly, based on
the use of an abstract base manifold, the matter space, $X$
say. This space (which in our case is three dimensional and
Riemannian) can be thought of as a book-keeping tool which assigns
unique labels to each fluid element via a map $\psimap$ that takes each
flow-line on spacetime ($M$ say) to a point in matter space. We may use
this map to push-forward the contravariant metric $g^{ab}$ on $M$ to $X$,
\beq
  g^{AB}=\psimap_* g^{ab},
\eeq
where abstract index notation \cite{wald:gr} with capital letters on
$X$ and lower case letters on $M$ is adopted. We then define the
tensor $\eta^{-1AB}$ to be the value of $g^{AB}$ that minimises the
energy density at a fixed number density $n$. We now define
$\eta_{AB}$ through the relation $\eta^{-1AC}\eta_{CB}=\delta^A{}_B$.
Pulling back $\eta_{AB}$ to $M$ ($\eta_{ab}=\psimap^*\eta_{AB}$)
we may define the constant volume strain tensor according to
\beq
  s_{ab} = \frac12(h_{ab}-\eta_{ab}),
\eeq
where $h_{ab}$ is defined by $h_{ab}=g_{ab}+u_au_b$ with $u^a$ being
the unit tangent vector to the flow-lines. Thus, in effect, $s_{ab}$
measures the difference in geometry between the natural, unsheared
state and the actual physical state.

In this paper we will use the simplifying assumption that the elastic
structure deforms conformally under pure compression. As discussed in
Paper I this allows us to fully describe this structure by a fixed
(\ie\ $n$-independent) metric tensor field $k_{AB}=n^{2/3}\eta_{AB}$
on $X$. One may think of this metric as measuring the relative
positions of the particles in a locally relaxed state.

Using the fact that the eigenvectors of $k_{ab}=\psimap^*k_{AB}$ are
eigenvectors also of the stress tensor $p_{ab}$ a preferred tetrad
completed by the matter four-velocity $u^a$ may be
introduced. Denoting the eigenvectors by $e_\mu^a$ we use Greek
indices to numerate the space-like basis vectors.  The eigenvalues
$n^2_\mu$ of $k_{ab}$ correspond, in a loose sense, to (squared)
linear particle densities, whereas the eigenvalues $p_\mu$ of the
strain tensor are the pressures as measured by a comoving observer in
the direction of $e_\mu^a$.

\subsection{Perturbation theory and notational remarks}

We shall assume that there exist a family of solutions to Einstein's
equations, parametrised by $\lambda$ say\footnote{We leave aside the
technical, but important, issue of existence of such a family. On
physical grounds, for the applications in mind in this paper, it is
apparent that such a family should exist.}. We assume that the
solution for a specific $\lambda$ (equal to zero say) to be known and
expand around this solution. We take $\delta$ in front of any tensor
field to represent the value of $\d/\d \lambda$ on that field
evaluated at $\lambda=0$. We shall have occasion to define perturbed
quantities without the explicit perturbation symbol $\delta$. The
indices on such quantities are raised and lowered by the unperturbed
spacetime metric, whereas care must be exercised when fields have retained
the $\delta$, \eg\
\beq
  \delta e_\mu^a = \delta (g^{ab}e_{\mu b}) \neq g^{ab}\delta e_{\mu b}.
\eeq
In order not to have too cluttered formulae we shall not notationally
distinguish background fields from the full family fields but instead
point out the validity of the equations in the text when confusion may
arise. 

In order to conform with the notation in the preceding papers in this
series\cite{ks:relasticityI,ksz:stability,ks:exact} as well as with
the work of Karlovini\cite{karlovini:axial} on which much of this paper is
based, we use some non-standard notation. In particular, the perturbed
metric will be denoted by $\gamma_{ab}$ and the flow-line orthogonal
piece of the metric is denoted by $h_{ab}$, \ie\ precisely the
opposite of the definitions in \eg\ \cite{carter:elastopert}.

We will use geometric units such that $c=1$, $G=1$ and the Einstein
equations take the form
\beq
  Z_{ab} := R_{ab} - \kappa(T_{ab}-\tsfrac12T^c{}_cg_{ab}) = 0
\eeq
where $R_{ab}$ is the Ricci tensor, $T_{ab}$ is the energy momentum
tensor and $\kappa$ is the coupling constant which, in conventional units,
takes the value $\kappa = 8\pi G c^{-4}$.

\section{General perturbations}\label{sec:general}

Perturbation theory of elastic media has already been considered by
Carter in some detail in \cite{carter:elastopert}. Since it seems
computationally advantageous to use the eigenvalue formulation in many
practical situations we shall devote this section to deriving the
perturbations of the eigenvectors and from that deducing the perturbed
stress energy tensor. The derivation is performed in an
arbitrary (identification) gauge and the final expression reduces to
the formulae of Carter when either a Lagrangian or a Eulerian gauge
(given by some displacement vector field $\xi^a$) is chosen. Since we are
assuming a perfectly elastic conformally deforming material (so we
have a fixed matter space metric $k_{AB}$) we shall consider the
perturbed matter space and spacetime metrics to be our fundamental
variables.

The perturbed eigenvectors are easily derived from the identities
\begin{displaymath}
 \begin{array}{ccc}
 g_{ab}\,u^a u^b = -1,  & 
 g_{ab}\,u^a e_\mu{}^b = 0, &
 g_{ab}\,e_\mu{}^a e_\nu{}^b = \delta_{\mu\nu}, \\
 k_{ab}\,u^a u^b = 0,  &
 k_{ab}\,u^a e_\mu{}^b = 0,  &
 k_{ab}\,e_\mu{}^a e_\nu{}^b = n_\mu^{\;2}\delta_{\mu\nu},
 \end{array}
\end{displaymath}
which, when perturbed, yield
\begin{align}
  u^a u^b\gamma_{ab} + 2u_a\,\delta u^a &= 0 \\
  u^a e_\mu{}^b\gamma_{ab} + e_{\mu\,a}\,\delta u^a + u_a\,\delta e_\mu{}^a &= 0 \\
  e_\mu{}^a e_\nu{}^b\gamma_{ab} + e_{\mu\,a}\,\delta e_\nu{}^a + 
  e_{\nu\,a}\,\delta e_\mu{}^a &= 0 \\ \lbeq{deltakeig}
  u^a u^b \delta k_{ab} &= 0 \\
  u^a e_\mu{}^b\delta k_{ab} + n_\mu^{\;2}e_{\mu\,a}\,\delta u^a &= 0 \\
  e_\mu{}^a e_\nu{}^b\delta k_{ab} + n_\mu^{\;2}e_{\mu\,a}\,\delta
  e_\nu{}^a + n_\nu^{\;2}e_{\nu\,a}\,\delta e_\mu{}^a &= 
  \delta(n_\mu^{\;2})\delta_{\mu\nu},
\end{align}
where $\gamma_{ab} = \delta g_{ab}$. Solving for $\delta u^a$, $\delta
e_\mu{}^a$ and $\delta(n_\mu^{\;2})$ gives
\begin{align}
  \delta u^a &= \tsfrac12 u^b u^c\gamma_{bc}\,u^a - \sum_{\mu=1}^3
      n_\mu^{\;-2}\,u^b e_\mu{}^c\delta k_{bc}\,e_\mu{}^a \\ 
  \delta e_\mu{}^a &= -\frac1{n_\mu^{\;2}}u^b e_\mu{}^c(\delta k_{bc}-n_\mu^{\;2}
      \gamma_{ab})\,u^a - \tsfrac12\,e_\mu{}^b e_\mu{}^c\gamma_{bc}\,e_\mu{}^a + \sum_{\nu\neq\mu}
      \frac1{n_\mu^{\;2}-n_\nu^{\;2}}\,e_\mu{}^b\e_\nu{}^c(\delta k_{bc} 
      - n_\mu^{\;2}\gamma_{bc})\,e_\nu{}^a \lbeq{deup} \\ 
  \delta(n_\mu^{\;2}) &= e_\mu{}^a e_\mu{}^b(\delta k_{ab} - n_\mu^{\;2}\gamma_{ab}). \lbeq{dn}
\end{align}
The perturbed basis one-forms are similarly found to be,
\begin{align}
  \delta u_a &=  -\tsfrac12 u^bu^c\gamma_{bc} u_a 
       - \sum_{\mu=1}^3 n_\mu^{-2}u^be_\mu^c(\delta k_{bc}-n_\mu^2\gamma_{bc})e_{\mu a} \\
  \delta e_{\mu a} &= -n_\mu^{-2}u^be_\mu^c\delta k_{bc} u_a 
       + \tsfrac12 e_\mu^be_\mu^c\gamma_{bc} e_{\mu a} - \sum_{\nu\neq\mu} \frac1{n_\nu^2-n_\mu^2} 
       e_\nu^be_\mu^c(\delta k_{bc} - n_\nu^2\gamma_{bc}) e_{\nu a} \lbeq{dedn}
\end{align}
One may be worried that the difference between the linear number
densities that appear in the denominator in the expressions
\refeq{deup} and \refeq{dedn} will cause divergences when there are 
degenerate eigenvalues on the background. However, as we will see, for
the physical quantities relevant for this work, the sums are always
convergent. These relations makes it a simple (but tedious) task to
write down the perturbations of any tensor field in terms of the
perturbed metrics. In particular the general perturbation of the
stress energy tensor takes the form
\begin{align}
  \delta T_{ab} &= (\delta\rho-\rho u^cu^d\gamma_{cd})u_au_b 
     + \sum_{\mu=1}^3(\delta p_\mu + p_\mu e_\mu^c e_\mu^d\gamma_{cd})e_{\mu a} e_{\mu b} \nonumber 
     - 2\sum_{\mu=1}^3u^ce_\mu^d[n_\mu^{-2}(\rho+p_\mu)\delta k_{cd} - \rho\gamma_{cd}]u_{(a}e_{\mu b)} \\
     &+ \sum_{\mu=1}^3\sum_{\nu\neq\mu}^3e_\mu^ce_\nu^d\left[n_\nu^{-2}(\rho + p_\nu)v_{\mu\perp\nu}^2(\delta k_{cd}
      -n_\nu^2\gamma_{cd}) +p_\nu\gamma_{cd}\right] e_{\mu (a}e_{\nu b)} \lbeq{dtab} 
\end{align}
where $v_{\mu\perp\nu}$ is the speed of shear waves along the
$\mu$-direction with polarisation vector along the $\nu$-direction, 
\beq\lbeq{swspeed}
(\rho+p_\nu)v_{\mu\perp\nu}^{\;2} =  
  \frac{n_\nu^{\;2}(p_\nu-p_\mu)}{n_\nu^{\;2}-n_\mu^{\;2}} 
\eeq
as derived in Paper I. It is now apparent that the isotropic limit is
well defined.  The perturbed energy density and eigenpressures can
also be expressed in terms of the perturbed metrics by the use of
\refeq{dn} (remembering that the equation of state is considered to be
given as a function of the linear particle densities) as
\begin{align}
 \delta\rho &= \frac12 \sum_\mu (\rho+p_\mu)n_\mu^{-2}(\delta k_{ab}-n_\mu^2\gamma_{ab})e_\mu^a e_\mu^b \\
 \delta p_\mu &= \frac12 \sum_\nu n_\nu\frac{\dell p_\mu}{\dell n_\nu}n_\nu^{-2}
        (\delta k_{ab}-n_\nu^2\gamma_{ab})e_\nu^a e_\nu^b \\
\end{align}
However, these will be seen to vanish for the axial case which is the
main concern of this paper.  One can confirm that, by comparing these
relations to the expression for the elasticity tensor given in Paper
I, one obtains the expression given by Carter
\cite{carter:elastopert}, a calculation that is best done in a Lagrangian 
gauge for which $u^a\delta k_{ab}$ vanish.

\section{Axial perturbations}

We shall now specialise our considerations to axial (in a sense to be
defined) perturbations. We use the approach of Karlovini
\cite{karlovini:axial}. For convenience, and to introduce notation, we 
briefly review this formalism below and then proceed to specialise it
to perfectly elastic matter.

Consider a spacetime with a Killing vector field $\eta^a$. If this
Killing vector field is nearly hyper-surface orthogonal, \ie\ its
twist
\beq
  \Omega^a = - \epsilon^{abcd}\eta_b\nabla_c\eta_d
\eeq 
can be considered to be small, we may treat the deviation from
hyper-surface orthogonality as a perturbation. Note that this general
considerations allows, as examples, studies of axisymmetric almost
spherical systems as well as stationary almost static systems (with no
required symmetry on the space-like hyper-surfaces). The main focus in
this paper is the spherical case, so that $\eta^a$ is the axisymmetry
generator and reduces to one of the SO(3) generators on the
background. In order to make progress we split the metric according to
\beq
  g_{ab} = \,\perp_{ab}\! + F\mu_a\mu_b,
\eeq
where
\beq
  F=\eta^a\eta_a, \qquad \mu_a=F^{-1}\eta_a, \qquad \eta^a\!\! \perp_{ab}=0
\eeq
On a spherical background the function squared norm of the Killing
vector has the form $F=(r\sin\theta)^2$. The polar and axial
perturbations of the spacetime metric\footnote{Note that the terms
'axial' and 'polar' will be inappropriate for perturbations of
non-spherical spacetimes.} are given by
\begin{align}
  {}^+\gamma_{ab} &= (\perp_a\!\!{}^c\!\!\perp_b\!{}^d\! + \mu_a\mu_b\eta^c\eta^c)\gamma_{cd} 
     = \delta\!\!\perp_{ab} + (\delta F)\mu_a\mu_b \lbeq{poldef} \\
  {}^-\gamma_{ab} &= 2\eta^c\mu_{(a}\!\!\perp_{b)}\!\!{}^d\gamma_{cd} = 2\eta_{(a}\delta \mu_{b)}. \lbeq{axidef}
\end{align}
Taking the appropriate projections of Einstein's equations,
\begin{align}
  \perp^{ac}\perp^{bd} \!\!Z_{ab} &= 0 \lbeq{eoo} \\
  \eta^a\eta^bZ_{ab} &= 0 \lbeq{eot}\\
  \perp^{ab}\!\eta^cZ_{ab} &= 0 \lbeq{ett}
\end{align}
and using a partial identification gauge fixing amounting to
\beq
 \delta\eta^a=0
\eeq
it is possible to show \cite{karlovini:axial} that, when evaluated at
$\Omega^a=0$, the axial part of the metric \refeq{axidef} satisfies
\refeq{eoo} and \refeq{eot} identically, whereas the polar part
\refeq{poldef} satisfies
\refeq{ett} identically. Hence these perturbations decouple and may be 
treated separately.

We henceforth restrict ourselves to the axial case whence
\beq
  \delta\!\!\perp_{ab}=0, \quad \delta F=0,
\eeq
so that
\beq
  \gamma_{ab}={}^-\gamma_{ab}=2\eta_{(a}\delta\mu_{b)}
\eeq
We next introduce the notation
\beq
  Q_{ab} = 2\nabla_{[a}\delta\mu_{b]}, \quad 
  J^a = 2\delta(\perp^{ab}\!\eta^cT_{bc})
\eeq
as well as their (restricted) duals
\beq
    Q^a = -\tsfrac12F\epsilon^{abcd}\eta_bQ_{cd},  
    \quad J_{ab} = \epsilon_{abcd}J^c\eta^d
\eeq
Note that both the perturbation vectors $Q^a$ and $J^a$ (as well as
their duals) are gauge invariant since the corresponding background
quantities vanish. In terms of these fields the full perturbation
equations takes the Maxwell-like form \cite{karlovini:axial}
\begin{align} 
  \nabla_{\!b}(FQ^{ab}) &= \kappa J^a, \lbeq{peq1} \\ 
  \nabla_{\!a}J^a &= 0, \lbeq{peq2}
\end{align} 
or in the dual form
\begin{align} 
  2\nabla_{[a}Q_{b]} &= -\kappa J_{ab}, \lbeq{dual1} \\ 
  \nabla_{\!a}(F^{-2}Q^a) &= 0, \lbeq{dual2} \\ 
  \nabla_{[a}J_{bc]} &= 0. \lbeq{dual3}
\end{align} 

To proceed it is evident that we must compute the axial matter current
vector $J^a$. To this end we assume that one of
the eigenvectors of the matter space metric, $e_3^a$ say, is
aligned with the axisymmetry generator on the background, \ie\
\beq
  e_3{}^a = F^{-1/2}\eta^a.
\eeq
It is worth pointing out that we hereby restrict ourselves to the case
when the Killing vector is space-like, but that no loss of generality
is implied in the spherically symmetric background case. Up to this
point the discussion has been valid (with appropriate changes of
terminology) to background spacetimes admitting any hyper-surface
orthogonal Killing vector.

To find $J^a$ we project \refeq{dtab} with $\perp^{ab}\!\!\eta^c$ and use
the relation
\beq
  \delta\!\!\perp^{\!ab} = - g^{ac}g^{bd}\gamma_{cd},
\eeq
so that, after some algebra, 
\begin{align}
  \eta^c T_{bc} \delta\!\!\perp^{\!ab} &= -p_3\eta^c\!\perp^{ab}\!\gamma_{bc}, \\
  \perp^{ab}\!\!\eta^c\delta T_{bc}    &= n_3^{-2}(\rho+p_3)\left(-u^au^b 
   + \sum_{\mu=1}^2e_\mu^a e_\mu^b v_{\mu\perp 3}^2\right)  
   (\delta k_{bc}-n_3^2\gamma_{bc}) + p_3\eta^c\!\perp^{ab}\!\gamma_{bc} \\
  &= (\rho+p_3)F\shm^{ab}K_b - \eta^c T_{bc} \delta\!\!\perp^{\!ab},
\end{align}
where $\shm^{ab}$ is the ``metric'' 
\begin{equation} 
  \shm^{ab} = -u^a u^b + \sum_{\mu=1}^2 v_{\mu\perp 3}^{\;2}\,e_\mu{}^a e_\mu{}^b, 
\end{equation} 
and $K_a$ is the gauge invariant (vanishing when unperturbed) 
quantity 
\begin{equation} 
  K_a = (F n_3^{\;2})^{-1}\delta(\perp_a\!{}^b\,\eta^c k_{bc}) =  
  (Fn_3^{\;2})^{-1}\perp_a\!{}^b\,\eta^c(\delta 
  k_{bc} - n_3^{\;2}\gamma_{bc}) = (F n_3^{\;2})^{-1}\perp_a\!{}^b\,\eta^c\delta 
  k_{bc} - \delta\mu_a. 
\end{equation}
Combining terms we finally find, 
\begin{equation} 
  J^a = 2(\rho + p_3)F \shm^{ab}K_b. \lbeq{JofK}
\end{equation} 
The tensor $\shm^{ab}$ is vaguely analogous to the so called acoustic metric
(sometimes denoted $G_{ab}$) used in \eg\ the study of analogues of
black holes in fluid mechanics in the sense that it is related to the
propagation of waves (although, in this case, it is shear waves rather
than sound waves). 

We still need to evaluate $K_a$. We therefore make the very natural
choice that the matter space is axisymmetric\footnote{One could, in
principle, consider a non-axisymmetric matter space and constrain the
mapping in such a way that spacetime is still axisymmetric. This
feels highly unnatural, however, and we will not consider it further.} 
and introduce material space coordinates $(\tilde{x}^i,\tilde\phi)$,
$i=1,2$, such that
\begin{equation} 
  \eta^a\nabla_{\!a}\tilde{x}^i = 0, \quad \eta^a\nabla_{\!a}\tilde\phi = 1.  
\end{equation} 
We also assume that the pull-back of the material space metric takes 
the form 
\begin{equation} 
  k_{ab} = \sum_{i,j}K_{ij}\nabla_{\!a}\tilde{x}^i\nabla_{\!b}\tilde{x}^j +  
  \tilde{F}\nabla_{\!a}\tilde\phi\nabla_{\!b}\tilde\phi, 
\end{equation} 
where the metric components $K_{ij}$ and $\tilde{F}$ depend on
$\tilde{x}^i$ only. Contracting this relation twice with $e_3^a$ we
find that the function $\tilde{F}$ may be expressed as
\beq
  \tilde{F}= F e_3^ae_3^bk_{ab} = F n_3^2,
\eeq
which holds on the background. 
This allows us to evaluate $K_a$ to be
\begin{equation}\lbeq{K} 
  K_a = \nabla_{\!a}\delta\tilde\phi-\delta\mu_a.
\end{equation} 
 
We may note in passing that an equivalent form of the
perturbation equations may be found directly in terms of the gauge
invariant one-form $K_a$. The equations of motion are then
\begin{equation} 
  \nabla_{\!b}(F\nabla^{[b}K^{a]}) = \kappa(\rho+p_3)F\shm^{ab}K_b,
\end{equation}
together with $\nabla_aJ^a=0$ with $J^a$ given in terms of $K_a$ by
\refeq{JofK}. We shall, however, not pursue this form further.

\subsection{Spherically symmetric background}

We shall now specialise the equations further by assuming that the
background is spherically symmetric. To ease the presentation we start
by introducing some notation. As much as possible we shall use the
notation of Paper II \cite{ksz:stability} in this series. The
background metric will be decomposed as,
\beq
 g_{ab} = j_{ab} + t_{ab},
\eeq
where $t_{ab}$ is the metric on the hyper-surfaces spanned by the
SO(3)-generators and is taken to be represented by the line element
\beq
  \d s_t^2 = r^2(\d\theta^2+\sin^2\!\theta\,\d\phi^2),
\eeq
where $r$ is the Schwarzschild radius. Introducing the shorthand
notation $r^a=e_1^a$, the orthogonal two dimensional Lorentzian metric
is given by
\beq
  j_{ab} = -u_au_b + r_ar_b,
\eeq
and the associated volume form is
\beq
  \epsilon_{ab} = -2u_{[a}r_{b]}
\eeq
The covariant derivative operator associated with $j_{ab}$ will be
denoted by ${\D}_a$.  Since the angular background eigenvectors
will now correspond to degenerate eigenvalues we may further introduce
the notation for the basis indices,
\beq
  \mu=1 \rightarrow \mu=r,\qquad \mu\in\{2,3\}\rightarrow \mu=t,
\eeq
that is, for example
\beq
  p_1=p_r, \qquad p_2=p_3=p_t.
\eeq

We now proceed to separate out the angular dependence from our
perturbations equations. The discussion will be very brief and the
reader is again referred to \cite{karlovini:axial} for details.

Using the dual form of the perturbation equations it is evident from
the closedness of $J_{ab}$ expressed in \refeq{dual3} that we may
introduce the vector potential $Y_a$ according to
\beq
  J_{ab} = 2\nabla_{[a}Y_{b]}.  
\eeq
Furthermore, this vector can be taken to be orthogonal to the
SO(3)-generators. Proceeding to equation \refeq{dual1} it is seen that
\begin{align} 
  Q_a = \nabla_{\!a}\Phi - \kappa Y_a 
\end{align} 
for some axisymmetric scalar $\Phi$. Separation is achieved by
putting
\begin{equation} 
  \Phi = C(\theta)r\psivar, \quad Y_a = C(\theta)\epsilon_{ab}{\Jcal}^b. 
\end{equation} 
The vector ${\Jcal}^a$ is a two-dimensional object related to $J^a$ by 
\begin{align}\lbeq{J} 
  J^a &= (r^2\sin\theta)^{-1}[\,C'(\theta){\Jcal}^a -  
    C(\theta){\Jcal}(\partial/\partial\theta)^a], 
\end{align} 
for the two-scalar ${\Jcal}$ constrained by \refeq{peq2} to take
the value ${\Jcal}={\D}_a{\Jcal}^a$. It is convenient
to introduce the invariantly defined mass function $m$ through
\beq
  1-\frac{2m}{r}={\D}_ar{\D}^ar,
\eeq
as well as the function $\tau$ invariantly defined on spherically
symmetric spacetimes by
\beq
  R^a{}_b\eta^b = \tsfrac12\kappa\tau \eta^a,
\eeq 
\ie\ $\frac12\kappa\tau$ is the eigenvalue of the Ricci tensor with 
respect to the eigenvector $\eta^a$. A clearer physical interpretation
is seen from the background Einstein equations which implies that
$\tau$ is the minus the trace of the $2\times 2$ block of the energy
momentum tensor orthogonal to the SO(3)-orbits. In the present case
that is
\beq
  \tau=\rho - p_r
\eeq
Noting also the relation
\beq
  r{\D}^a{\D}_a r = \frac{2m}{r}-\frac12\kappa r^2\tau
\eeq
the perturbation equations can straight forwardly be confirmed to
reduce to the two-dimensional equations
\begin{align} 
  ({\D}_a{\D}^a - U)\psivar = \kappa S \lbeq{Peqsep1} \\ 
  {\D}_a{\Jcal}^a = {\Jcal}, \lbeq{Peqsep2}
\end{align} 
where 
\begin{equation} 
  S = r\epsilon^{ab}{\D}_a(r^{-2}{\Jcal}_b),
\end{equation} 
and the potential $U$ is given by
\beq
  U = \frac12\kappa\tau - \frac{6m}{r^3} + \frac{l(l+1)}{r^2},
\eeq
where $l$ is the usual harmonic index coming from the separation of
variables. The angular equation is solved by setting
\begin{align}
  C(\theta) &= G_{l+2}^{-3/2}(\cos\theta)
\end{align}
where $G_{l+2}^{-3/2}(y)$ is an ultra-spherical (or Gegenbauer) polynomial.

In our case $J^a$ is given in terms of background fields and $K_a$, 
where, as implied by eq.\ \refeq{K},  
\begin{equation} 
  2\nabla_{[a}K_{b]} = -2\nabla_{[a}\delta\mu_{b]} = -Q_{ab}. 
\end{equation}
This means that $\delta\mu_a$ must be of the form 
\begin{align} 
  \delta\mu_a = -\frac{C'(\theta)}{(l+2)(l-1)r^2\sin^3\!\theta}\epsilon_a{}^b 
  \Qcal_b + \nabla_a \fgauge \qquad &(l\ge2) \\
  \delta\mu_a = r^{-2}\epsilon_a{}^b\Qcal_b + \nabla_a \fgauge \qquad &(l=1)
\end{align}
where
\beq
  \Qcal_a = {\D}_a(r\psivar)-\kappa\epsilon_{ab}{\Jcal}^b, \lbeq{Qdef}
\eeq
and $\fgauge$ is a free function (which we will see in section
\ref{sec:strain} correspond to the choice of gauge).  This in turn
implies that $K_a$ must be of the form
\begin{align} 
  K_a = \frac{C'(\theta)}{(l+2)(l-1)r^2\sin^3\!\theta} 
  \epsilon_a{}^b\mathcal{Q}_b + \frac1{(l+2)(l-1)}\nabla_{\!a}[\sin^{\!-3}\!\theta\,C'(\theta)r^{-1}\varphi]  &\nonumber \\ 
  = \frac{C'(\theta)}{(l+2)(l-1)r^2\sin^3\!\theta} 
  [\epsilon_a{}^b\mathcal{Q}_b + r^2{\D}_a(r^{-1}\varphi)]  
  - \frac{C(\theta)}{\sin^3\!\theta}\,r^{-1}\varphi\nabla_{\!a}\theta,  &\qquad (l\ge2) \lbeq{Kid} \\
  K_a = r^{-2}\epsilon_a{}^b\Qcal_b + \D_a(r^{-1}\varphi) &\qquad (l=1)
\end{align} 
for some spherically symmetric scalar $\varphi$. Setting 
\begin{equation} 
  \shmtwoD^{ab} = -u^a u^b +{\vr}r^a r^b,  
\end{equation} 
we have 
\begin{equation} 
  \shm^{ab} = \shmtwoD^{ab} + v_{t\perp t}^{\;2}e_2^{\;a}e_2^{\;b}.
\end{equation} 
For $l\ge2$ we then find 
\begin{align} 
  J^a &= 2(\rho+p_t)F\shm^{ab}K_b \nonumber \\ 
  &= \frac{2(\rho+p_t)}{\sin\theta} 
 \left\{ \frac{C'(\theta)}{(l+2)(l-1)}  
 \shmtwoD^{ab}[
 \epsilon_a{}^b\mathcal{Q}_b + 
 r^2{\D}_a(r^{-1}\varphi)] -  
 C(\theta)v_{t\perp t}^{\;2}\,r^{-1}\varphi\,(\partial/\partial\theta)^a \right\} 
\end{align} 
Comparing with eq.\ \refeq{J}, we obtain
\begin{align} 
  [\Ecal\shmtwoD^a{}_b+(l+2)(l-1)j^a{}_b]\kappa{\Jcal}^b  
  &= \Ecal\shmtwoD^{ab} 
  [\epsilon_b{}^c{\D}_c(r\psivar)+r^2{\D}_b(r^{-1}\varphi)] \\ 
  \kappa{\Jcal} &= \Ecal v_{t\perp t}^{\;2}r^{-1}\varphi \lbeq{Jsol}
\end{align}
where
\beq
  \Ecal = 2\kappa r^2(\rho+p_t).
\eeq 
Solving for $\kappa{\Jcal}^a$ we find
\begin{equation}
  \kappa{\Jcal}^a = (-C_0u^a u^b + C_1r^a r^b) \lbeq{Jasol}
  [\epsilon_b{}^c{\D}_c(r\psivar)+r^2{\D}_b(r^{-1}\varphi)]  
\end{equation} 
where 
\begin{align} 
  C_0 &= \frac{\Ecal}{\Ecal+(l+2)(l-1)} \lbeq{C0}\\ 
  C_1 &= \frac{\Ecal{\vr}}{\Ecal{\vr} 
  +(l+2)(l-1)}. \lbeq{C1} 
\end{align} 
When $l=1$ a similar analysis shows that $\Jcal=0$ and that $\Jcal^a$
is given by the expression \refeq{Jasol} with 
\begin{align} 
  C_0 &= \frac{\Ecal}{\Ecal+1} \\ 
  C_1 &= \frac{\Ecal{\vr}}{\Ecal{\vr}+1}. 
\end{align} 

Finally, rewriting the perturbation equations
\refeq{Peqsep1}-\refeq{Peqsep2} in terms of the quantity
$\mathcal{Q}_a$ defined in \refeq{Qdef}, we find
\begin{align} 
  &r\,{\D}^a(r^{-2}\mathcal{Q}_a) 
   = \frac{(l+2)(l-1)}{r^2}\psivar, \lbeq{PeqQ}\\ 
  &{\D}_a{\Jcal}^a = {\Jcal}, \lbeq{PeqJ}  
\end{align} 
where $\mathcal{Q}_a$, ${\Jcal}^a$ and ${\Jcal}$ are given in
terms of the two-dimensional scalar fields $\psivar$ and $\varphi$ in eqs.\
\refeq{Qdef}, \refeq{Jasol} and \refeq{Jsol} respectively. These equations are valid for  
any $l\ge 1$ provided one remember to put $\Jcal=0$ when $l=1$. The
non-radiative nature of the $l=1$ case is hinted by the form of the
equations which are just decoupled conservation equations for the two
``currents'' $r^{-2}\Qcal^{\,a}$ and $\Jcal^a$.

In order to better understand the content of these equations it is
useful to write them out in a suitable coordinate system. For brevity
we will from now on only consider the radiative case $l\ge2$. It is
advantageous to use explicitly conformally flat coordinates given by
\beq 
  u_a = -e^{\nu}(\d t)_a, \qquad r_a = e^{\nu}(\d\rstar)_a.
\eeq
which will reduce to the usual Regge-Wheeler radial gauge on static
backgrounds. Next we introduce the notation
\beq
  \mathcal{X}_t = u_a\mathcal{X}^a, \qquad \mathcal{X}_r = r_a\mathcal{X}^a
\eeq
for any $\mathcal{X}^a$. Then we have
\begin{align}
  \kappa\Jcal_t &= e^{-\nu}C_0\left[\Ap + r^2\Bdot\right] \lbeq{Jt} \\
  \kappa\Jcal_r &= e^{-\nu}C_1\left[\Adot + r^2\Bp\right] \lbeq{Jr} \\
  \Qcal_t &= e^{-\nu}\left[(1-C_1)\Adot - C_1 r^2\Bp\right] \lbeq{Qt} \\
  \Qcal_r &= e^{-\nu}\left[(1-C_0)\Ap - C_0 r^2\Bdot\right] \lbeq{Qr} \\
\end{align}
where dots and primes denotes derivatives with respect to $t$ and
$\rstar$ respectively\footnote{This notation should not be confused
with the use of dots and primes in Paper II \cite{ksz:stability}.}. Using
\beq
  \Wcal_t=r^{-1}e^\nu\Qcal_t, \qquad \Wcal_r=r^{-1}e^\nu\Qcal_r
\eeq
as auxiliary variables we can cast the perturbation equations as a
first order system,
\begin{align}
  -\dot\Wcal_t + \Wcal_r' + \frac{\dot{r}}{r}\Wcal_t -\frac{r'}{r}\Wcal_r - e^{2\nu}\frac{L}{r^2}\psivar &= 0 \lbeq{1ogen1}\\
  -\dot{\Wcal}_r + \Wcal_t' - \frac{\dot{r}}{r}\Wcal_r +\frac{r'}{r}\Wcal_t + e^{2\nu}\frac{\Ecal\vt}{r^2} \varphi &= 0 \lbeq{1ogen2}\\
  -L\Adot + \Ecal\vr r^2\Bp + (\Ecal\vr +L)r\Wcal_t &= 0  \lbeq{1ogen3}\\
  L\Ap - \Ecal r^2\Bdot - (\Ecal+L)r\Wcal_r &= 0 \lbeq{1ogen4}
\end{align}
where $L=(l+2)(l-1)$ and we have used the definitions of $C_0$ \refeq{C0} and $C_1$
\refeq{C1}. We may note here that the principal part of these equations 
decouple into two systems, one in $\psivar$ and $\varphi$ which has a
characteristic propagation speed $\vr$ and one in $\Wcal_t$ and
$\Wcal_r$ whose speed is $1$. We may naturally interpret this as the
existence of two families of modes, the shear modes and the axial
gravitational $w$-modes.

This is as far as we go in the general case. Note that this system is
suitable for numerical integration since it does not contain
derivatives of background quantities that can be discontinuous. As we
will see below, it also makes maximal use of the junction conditions
so that $\Wcal_t$, $\Wcal_r$ and $\psivar$ are all everywhere continuous.

\section{Static background}

The above derived system \refeq{1ogen1} - \refeq{1ogen4} of equations
is valid for any spherically symmetric background and could therefore
be applied to \eg\ collapsing bodies. However since our interest here
lies on matter with non-zero shear modulus such configurations are not
realistic (since the strain would inevitably grow beyond the breaking
strain of the material). One could of course study axial perturbations
of a radially oscillating star, but henceforth we will restrict our
attention to the case where the background is static. We can then drop
all dotted background quantities. It is then possible to combine
\refeq{1ogen1} - \refeq{1ogen3} to give a wave equation for $\Wcal_t$ 
sourced by $\varphi$. In addition the equations \refeq{1ogen2} -
\refeq{1ogen4} can be combined into a wave equation for $\varphi$ with 
$\Wcal_t$ as a source;
\begin{align}
  -\ddot\Wcal_t + \Wcal_t'' + e^{2\nu}\left(\frac{6m}{r^3} - \frac12\kappa(\rho-p_r)\right.
      & \left. -\frac{\Ecal\vr}{r^2} - \frac{l(l+1)}{r^2}\right)\Wcal_t \nonumber\\ 
      &= r\left[\frac{e^{2\nu}\Ecal(\vr-\vt)}{r^3}\varphi\right]' - \left[\frac{e^{2\nu}\Ecal\vr}{r^2}\right]'\varphi \lbeq{Weq1}\\
  -\ddot\varphi + \frac{(\Ecal\vr\varphi')'}{\Ecal}  - \left[\frac{(\Ecal\vr r')'}{\Ecal r}\right. & \left.+ 
      e^{2\nu}\frac{\vt(\Ecal+L)}{r^2}\right]\varphi 
      = \frac{[\Ecal(1-\vr)r\Wcal_t]'}{\Ecal r} - \frac{\Ecal'}{\Ecal} \Wcal_t\lbeq{Weq2}
\end{align}
Of course, many possible reformulations exist of this system of
equations if new dependent variables are chosen. One such
reformulation would have been to choose $\psi$ instead of $\Wcal_t$ as
the first dependent variable. This would have followed the general
guidelines in \cite{karlovini:axial} more closely, but would not have
lead to simplified formulae in this case. We do not feel that any
possible second order system reformulation (that we have found) is
superior in any important ways to  \refeq{Weq1}-\refeq{Weq2}. In
any case, for numerical integration it is preferable to use the first
order system
\refeq{1ogen1}-\refeq{1ogen4}. We could also give more explicit
expressions of the terms involving derivatives of background
quantities by making use of the results of Paper I.  However, either
one then keeps the elastic equation of state free in which case third
order partial derivatives of the energy per particle will appear, or
one specify \eg\ a quasi-Hookean equation of state leading to rather
lengthy expressions. We do not feel that this leads to any further
understanding of the problem at hand. Instead we give explicit
expressions below suitable for numerical integration.  We shall
therefore be satisfied with this form of the wave equations and pause
a moment to discuss some of their properties.

First, it is clear that for vacuum \refeq{Weq2} is trivial and
\refeq{Weq1} is just the usual Regge-Wheeler equation. For this 
reason we may refer to \refeq{Weq1} as the gravitational wave
equation. Second, if we take the isotropic limit we have
$\vr=\vt=2\kappa r^2\muc\Ecal^{-1}$, where $\muc$ is the shear
modulus, so that the source terms in
\refeq{Weq1} reduce to
\beq
  2\kappa (e^{2\nu} \muc)'\varphi
\eeq
so that we see that, as pointed out by Dyson \cite{dyson:seismic},
for weak gravity, $e^{\nu}\approx 1\approx$ constant, gravitational
waves couple only to the gradient of the shear modulus. Another
important case is the perfect fluid limit. Here we see that the system
reduces to 
\begin{align}
  &-\ddot\Wcal_t + \Wcal_t'' + e^{2\nu}\left(\frac{6m}{r^3} - \frac12\kappa(\rho-p) - \frac{l(l+1)}{r^2}\right)\Wcal_t = 0 \\
  &-r\varphi^{\lagomdot\lagomdot} = (r\Wcal_t)' \lbeq{pfmat}
\end{align}
The first of these equations is the standard result, see \eg\ 
\cite{cf:osc,cf:gwresonance,kokkotas:axialmodes}. The second
equation is now redundant since the variable $\varphi$ does not encode
any relevant physical information in this case\footnote{The extra
equation correspond in a sense to the freedom of choice of mapping
between $M$ and $X$. For perfect fluids the labelling of fluid
elements plays no role in the dynamics so there is some degree of
degeneracy in defining the map.}. It is in fact easy to show, using
\refeq{Jt}-\refeq{Qt} and \refeq{pfmat}, that the two dimensional
matter current $\Jcal^a$ satisfy $\Jcal_r=0$, $\dot{\Jcal}_t=0$ so
that the well known result that axial perturbations of perfect fluids
reduce to stationary currents is evident.

\subsection{Boundary conditions}
The perturbation equations must be accompanied by suitable boundary
conditions. These conditions were treated in detail by Karlovini for
general matter sources \cite{karlovini:axial} and we merely state the
results here; At any boundary (including \eg\ interfaces between
different layers or the surface) we require that the one-form
$\mathcal{Q}_a$ as well as the scalar $\psivar$ should be
continuous. These conditions follow from the requirement that the
first and second fundamental form on any hyper-surface with normal
$n^a$ (say) are continuous. In particular this implies that the
traction $n^aT_{ab}$ is continuous as well \cite{ms:junction}. 
In ST the condition of continuous traction was enforced as an
additional constraint. In view of the above it is clear that this is
only necessary as a gauge condition relating the gauges on either
side of the hyper-surface. An additional boundary condition is provided
by requiring regularity at the centre. Finally, if we consider an
isolated object, we need to impose a condition of outgoing
gravitational waves at infinity. We discuss these boundary conditions
in more detail in the next section where we give a computational recipe
for quasi normal mode solutions to the perturbation equations.

\section{Computational algorithm}
\label{sec:comp}

We shall here present a procedure for finding quasi-normal mode
solutions to the perturbation equations in situations relevant for
neutron stars with a fluid core, a solid crust and, possibly, a fluid
ocean. We therefore assume harmonic time dependence, \ie\ that the
temporal dependence of the independent variables is given by
$e^{i\omega t}$. We have found it slightly preferable to present the
equations in Schwarzschild coordinates, related to the Regge-Wheeler
coordinates through
\beq
  e^\nu\d\rstar = e^\lambda\d r
\eeq
where $\lambda$ is given by
\beq
  e^{-2\lambda} = 1-\frac{2m}{r}
\eeq
When we wish to solve our perturbation equations numerically it is
preferable that all variables scale with the same power of $r$ near
the centre. To obtain this behaviour (as well as some other features)
we redefine the variables in the system \refeq{1ogen1}-\refeq{1ogen4}
(specialised to a static background) according to
\begin{align}
  \Xcal_1 &= r^{-l-1}\Wcal_t\\
  \Xcal_2 &= -i\omega_0r^{-l}\Wcal_r \\
  \Xcal_3 &= -i\omega_0e^{\nu}r^{-l-1}\psivar \\
  \Xcal_4 &= \omega_0^2e^{\nu}r^{-l}\varphi
\end{align}
where $\omega_0 = e^{-\nu_c}\omega$ and $\nu_c$ is the central value
of $\nu$. The reason for introducing these variables is that if one
wish to solve the background equations simultaneously with the
perturbation equations using a shooting algorithm the central value of
$\nu$ is unknown since it is determined by matching the background
solution to the exterior Schwarzschild solution at the surface of the
star. The above scaling allows for writing the equations in terms of
$\omega_0$ and $\nu_0=\nu-\nu_c$ and, once the solution is found scale
it back to physical units. 

It is easy to show that for perfect fluids
the last two of the first order equations become algebraic relations
amounting to
\beq\lbeq{pfconst}
  \Xcal_3= -e^{\nu_0}\Xcal_1, \qquad \Xcal_4 = -e^{\nu_0}\Xcal_2
\eeq
whereas the other two become just
\begin{align}
  r\frac{\d\Xcal_1}{\d r} &= -(l+2)\Xcal_1 - e^{\lambda-\nu_0}\Xcal_2 \\
  r\frac{\d\Xcal_2}{\d r} &= -e^{\lambda-\nu_0}\left[Le^{2\nu_0} - \omega_0^2r^2\right]\Xcal_1 - (l-1)\Xcal_2
\end{align}
Disregarding the singular solution scaling as $r^{-2l-1}$ near
$r=0$ it is easy to see that the variables leave the centre according to
\begin{align}
  \Xcal_1 &= \hat\Xcal_1 + O(r^2) \lbeq{exp1}\\
  \Xcal_2 &= -(l+2)\hat\Xcal_1 + O(r^2)\\
  \Xcal_3 &= \hat\Xcal_1 + O(r^2)\\
  \Xcal_4 &= -(l+2)\hat\Xcal_1 + O(r^2) \lbeq{exp4}
\end{align}
where $\hat\Xcal_1$ is an arbitrary constant.
The equations in a solid becomes,
\begin{align}
  r\frac{\d\Xcal_1}{\d r} &= -(l+2)\Xcal_1 - e^{\lambda-\nu_0}\Xcal_2 - e^\lambda \frac{\Ecal\vt}{r^2\omega_0^2}\Xcal_4 \\
  r\frac{\d\Xcal_2}{\d r} &= e^{\lambda-\nu_0}\omega_0^2r^2\Xcal_1 - (l-1)\Xcal_2 + e^\lambda L\Xcal_3 \\
  r\frac{\d\Xcal_3}{\d r} &= e^\lambda \frac{\Ecal+L}{L}\Xcal_2 + [r\nu_{,r} - (l+2)]\Xcal_3 
        + e^{\lambda-\nu_0}\frac{\Ecal}{L}\Xcal_4 \\
  \Ecal\vr r\frac{\d\Xcal_4}{\d r} &= -e^{\lambda}r^2\omega_0^2[\Ecal\vr + L]\Xcal_1
        - e^{\lambda-\nu_0}r^2\omega_0^2L\Xcal_3 
        + \Ecal\vr[r\nu_{,r} - (l-1)]\Xcal_4  
\end{align}
where
\beq
  \nu_{,r} = \frac{\d \nu}{\d r} = \frac{m + \tsfrac12 \kappa r^3 p_r}{r(r-2m)}
\eeq
and the shear wave speeds depend on the equation of state according to
\refeq{swspeed}. For the quasi-Hookean equation of state discussed in Paper 
I they take the form 
\begin{align}
  \Ecal\vr &= \kappa\muc r^2\left[1+\frac{n_t^2}{n_r^2}\right] \\
  \Ecal\vt &= 2\kappa\muc r^2\left[\frac{n_t^2}{n_r^2} + \frac{n_r^2}{n_t^2} -1 \right]
\end{align}
In vacuum the equations are
\beq\lbeq{vacconstr}
  \Xcal_3= -e^{\nu_0}\Xcal_1, \qquad \Xcal_4 = \mbox{unconstrained}
\eeq
\begin{align}
  r\frac{\d\Xcal_1}{\d r} &= -(l+2)\Xcal_1 - \frac{re^{\nu_c}}{r-2M}\Xcal_2 \\
  r\frac{\d\Xcal_2}{\d r} &= -e^{-\nu_c}\left[L - \frac{e^{2\nu_c}\omega_0^2r^3}{r-2M}\right]\Xcal_1 - (l-1)\Xcal_2
\end{align}
where $M$ is the total mass of the star.

Using the expansion \refeq{exp1}-\refeq{exp4} we may now integrate the
fluid equations to the crust core interface, where we need to impose
the boundary conditions. In the present variables these are just the
continuity of $\Xcal_1$, $\Xcal_2$ and $\Xcal_3$. Note that $\Xcal_4$
is free at the boundary and can in principle be set to any
value. However, as we proceed with the integration in the solid we
eventually encounter the next boundary. If this boundary is an
interface to a fluid phase or vacuum we have two conditions on the
variable $\Xcal_3$, i) it must be continuous and ii) it has to satisfy
the constraint \refeq{pfconst} (which is identical to \refeq{vacconstr}
in the vacuum case). In general these will not be satisfied signalling
the wrong choice of value of $\Xcal_4$ at the previous boundary. The
remedy to this situation is to find a second independent solution to
the equation in the solid phase and then to take the linear
combination of the two solutions which satisfy all the boundary
conditions. The simplest way to find the second solution is to start
with $\Xcal_1=\Xcal_2=\Xcal_3=0$ and $\Xcal_4=$ any value. Clearly,
adding any multiple of this solution to the original one will not
violate the initial boundary conditions so it is a simple task of
solving an algebraic equation to find the correct total solution in
this part of the star. It should be evident how to work out the entire
solution to the problem using this algorithm regardless of the number
of interfaces in the star.
Finally, the quasi-normal mode frequencies $\omega$ are determined by
making sure that the vacuum solution is described by outgoing
gravitational waves. This specifies the solutions up to a scale.



\section{Identifying the metric and the strain} \label{sec:strain}
Since neither the metric nor the strain on a strained background
is gauge invariant we need to specify the gauge in order to evaluate
them. We have already used some of our gauge freedom by
setting $\delta \eta^a=0$, so any further gauge transformation has to
be generated by a vector field, $\zeta^a$ say, that is Lie-dragged by
the axisymmetry generator,
\beq
 \lie_\eta \zeta^a=-\lie_\zeta \eta^a = 0
\eeq
We may decompose $\zeta^a$ according to 
\beq
  \zeta^a = \zeta_\perp^a + \fgauge\eta^a, \quad \zeta_\perp^a\eta_a=0, \quad \eta^a\nabla_a \fgauge=0
\eeq
Now, since 
\beq
  \lie_{\zeta_\perp}\mu_a = 0
\eeq
due to the fact that $\eta^a$ is a hyper-surface orthogonal Killing
vector on the background it is clear that gauge transformations
generated by a vector field orthogonal to, and Lie-dragged by $\eta^a$
does not affect the axial perturbations. Likewise, it is easy to show
that a gauge transformation of the type $\zeta^a=\fgauge\eta^a$ does
not affect the polar perturbations. Hence we need only consider gauge
transformations of the latter type. Under such a transformation we
have
\begin{align}
  \delta\mu_a &\rightarrow \delta\mu_a + \nabla_a \fgauge, \\
  \delta\tilde\phi &\rightarrow \delta\tilde\phi + \fgauge.
\end{align}
Since $\eta^a\nabla_a \delta\tilde\phi=0$ we may use this freedom to
set $\delta\tilde\phi=0$\footnote{Note that Schumaker and Thorne uses
the gauge freedom to set the $\theta$-component of
$\delta\mu_a=0$. This would correspond to setting $\fgauge=$
constant. It is also worth pointing out that, since these authors
consider isotropic backgrounds, their perturbed strain tensor is gauge
invariant.}. This gauge choice is of a comoving, or Lagrangian type
and is the natural way to measure the strain since
\eg\ the breaking strain will invariably be calculated in a comoving frame. 
We thus find that in a Lagrangian gauge we have
$\Delta\tilde\phi=0$ so that $\fgauge=-\delta\tilde\phi$ and
$\Delta\mu_a = -K_a$, where we use $\Delta$ to indicate that a special
gauge has been chosen. Using these relations we find
\beq\lbeq{spert}
  \Delta s_{ab} = -\eta_{(a}K_{b)} + u_{(a}\Delta u_{b)} 
     = -\eta_{(a}h_{b)}{}^{c}K_c. 
\eeq
The perturbed metric is just
\beq \lbeq{gpert}
  \Delta g_{ab} = -2\eta_{(a}K_{b)}
\eeq
and
\beq
  \Delta k^a{}_b = 2k_{bc}\eta^{(c}K^{a)}.
\eeq

Usually estimates of the breaking, or yield, strain is given as a
dimensionless strain angle, $\Theta_{\brk}$ say. The precise relation to
the components of the strain tensor depends on the type of deformation
and the microscopic structure of the solid material. However, for
simple (\eg\ isotropic or cubic) structures under simple deformations
(\eg\ pure twist or shear) the nonzero components of $s_{ab}$ have the
form
\beq
  s_{ab} \approx \tsfrac12 \Theta.
\eeq
Given the uncertainty in the literature on the value of
$\Theta_{\brk}$ \cite{ruderman:tectonicsIII} it does not feel
meaningful to digress too deeply into the subject of breakdown of
elasticity. It is clear that the approximation of perfect elasticity
will break down before the material actually cracks, so for all
purposes of this paper we may assume that something catastrophic
happens when any component of $s_{ab}$ exceeds some value of the order
of $\frac12\Theta_{\brk}$.

An interesting observation is that, even though the first order
perturbations of the linear particle densities are all zero, so that
the strain scalar also is unperturbed to first order, we may still
estimate the second order contribution to the strain scalar and
thereby estimate the energy stored in the elastic material. We may
namely compute the strain scalar by first evaluating the total
\beq
 k^a{}_b = {}^0k^a{}_b + \Delta k^a{}_b + \Delta^2 k^a{}_b
\eeq
for some unspecified second order perturbation $\Delta^2
k^a{}_b$. Noting that the non-orthogonal pieces will not alter the
strain scalar (all such pieces are contracted with orthogonal tensors)
and using bold face letters to denote the $3\times 3$-matrices $\bldk
= h^a{}_ck^c{}_b$ we may write (see Paper I)
\beq
  s^2 = \frac1{36 \det \bldk}[(\Tr \bldk)^3 - \Tr(\bldk^3) - 24\det\bldk]
\eeq
It then turns out (after some straightforward but tedious algebra)
that the second order perturbation terms only comes in multiplied by
the background strain. These terms will therefore always (due to the
small numerical value of the breaking strain) be much smaller than
other second order terms. Neglecting them leads to an expression of
the form
\beq\lbeq{s2est}
  s^2 = s_0^2 + \frac12Fh^{ab}{K}_a{K}_b + \ldots 
\eeq
where $s_0$ is the background value.

In summary, once a solution to the perturbation equations are found
using the procedure described in section \ref{sec:comp} it is easy to
find the metric and stress tensor perturbations from equations
\refeq{gpert} and \refeq{spert} respectively. Since the solutions will 
only be defined up to a scale one may then set this scale such that
the largest component of $s_{ab}+\Delta s_{ab}$ is smaller that some
maximum value of order $\frac12\Theta_{\brk}$ to find the maximally allowed
amplitude of the perturbations consistent with elastic response. It is
also straight forward to estimate the maximal stored elastic energy
density by the formula
\beq
  \rho_\mrm{elastic} = \muc s^2
\eeq
where $\muc$ is the shear modulus and $s^2$ is given by \refeq{s2est}.

\section{Conclusion}

We have developed the general relativistic theory of torsional
oscillations in elastic matter. In the light of the recent very
exciting observations of quasi-periodic oscillations in the tail of
giant flares in soft gamma-ray repeaters, with frequencies matching
well the expected spectrum of such modes, we believe that it is
important to have a well founded theory for these modes. We have
improved on the previously existing theory (mainly ST) in several
respects; The equations presented here are gauge invariant and are
valid to second order in strain for any equation of state as long as
the material is of a conformally deforming type. Various gauge-issues
have been resolved and resulted in simplified equations and boundary
conditions compared to ST. The elastic response to perturbations is
conveniently parametrised in terms of the shear wave velocities and
should be straight forward to apply to the anisotropic ``pasta''
phases near the crust-core boundary of neutron stars once the relevant
elastic parameters have been estimated.

The results of this paper has in fact already been applied in the
relativistic Cowling approximation \cite{sa:axicowling}. Although it
is expected that the spectrum should not change substantially when
gravitational degrees of freedom are included it is nevertheless
important to check this numerically and work in that direction is in
progress.

\section*{Acknowledgements}
LS gratefully acknowledge support by a Marie Curie Intra-European
Fellowship, contract number MEIF-CT-2005-009366. Parts of this work
was carried out when the authors were at Stockholm University. We are
grateful to Nils Andersson and Brandon Carter for discussion and
insightful comments. In the course of this work extensive use was made
of the computer algebra package GRTensorII \cite{mpl:grtensorII}
running within Maple. We also acknowledge support from the EU-network
ILIAS providing opportunity for valuable discussions with our European
colleagues.

\end{document}